\documentclass[twocolumn,showpacs,preprintnumbers,amsmath,amssymb]{revtex4}

\usepackage{graphicx}
\usepackage{dcolumn}
\usepackage{bm}

\begin{document}

\preprint{APS/123-QED}

\title{Complete population transfer in 4-level system via Pythagorean triple coupling}

\author{Haim Suchowski$^{1,3}$}
\email{haim.suchowski@weizmann.ac.il}
\author{Dmitry B. Uskov$^{2,3}$}
\email{uskov@tulane.edu}

\affiliation{$^{1}$Department of Physics of Complex Systems,
Weizmann Institute of
Science, Rehovot 76100, Israel\\
$^{2}$Department of Physics and Engineering Physics, Tulane University, New
Orleans, LA 70118, and Hearne Institute for Theoretical Physics at Louisiana State University, Baton Rouge LA, 70806\\
$^{3}$Kavli Institute for Theoretical Physics, Santa Barbara, CA
93106, USA}


\date{\today}

\begin{abstract}
We describe a relation between the requirement of complete
population transfer in a four-mode system and the generating
function of Pythagorean triples from number theory. We show that
complete population transfer will occur if ratios between coupling
coefficients exactly match one of the Pythagorean triples $(a,b,c)
\subset Z$, $c^{2}=a^{2}+b^{2}$. For a four-level ladder system this
relation takes a simple form $\left( {V_{12} ,V_{23} ,V_{34} }
\right) \sim (c,b,a)$, where coefficients $V_{ij}$ describe the
coupling between modes. We find that the structure of the evolution
operator and the period of complete population transfer are
determined by two distinct frequencies. A combination of these
frequencies provides a generalization of the two-mode Rabi frequency
for a four-mode system.

\end{abstract}
 \maketitle

Revealing hidden mathematical structures behind physical phenomena
is of great importance, especially when qualitative dynamical
properties of a quantum system reduce to a basic relation from the
number theory. Here we show how the Pythagorean triple, which is the
set of three integer numbers $(a,b,c)$, satisfying the Pythagorean
equation $a^{2}+b^{2}=c^{2}$, is found to play a significant role in
the dynamics of a four-mode system.

Describing the evolution of a general multi-mode system and finding
conditions for complete  population transfer from one mode to
another is a subject of extensive research for a variety of
classical and quantum systems. Coherent manipulation of population
of states in atomic and molecular quantum systems \cite{Eberly1975},
spin control in nuclear magnetic resonances \cite{NMR2005}, quantum
information processing \cite{Nielsen2000, Sorenson1999}, and
directional optical waveguide technology \cite{Yariv1973} are only a
few examples where complete population transfer is desired. Here we
study the four-mode dynamics, which is of particular importance for
the quantum information processing technology, where two-qubit
quantum logic gates serve as elementary building
blocks for designing fully functional scalable devices \cite{Nielsen2000,Uskov2008,Rau2000}. 

 In general, solutions of dynamic coupled equations are
difficult to analyze, and even for the simplest case of a two-level system, realized by a spin-$\frac{1}{2}$ particle or a two-level atomic system, only a handful of analytical solutions are known in the literature.  
The simplest one, known as the Rabi solution \cite{Eberly1975},
describes a two-level system with a constant coupling. For this
solution, complete population transfer between two states occurs
only when the frequency of an external driving field is on resonance
with the energy difference between the modes. The period of complete
population transfer, called the Rabi flopping time, is inversely
proportional to the
strength of the mode coupling. 
 A geometric visualization of a two-level system by representing its state by, a point within the Bloch
 sphere, plays important role in developing a clear intuitive understanding of the two-mode dynamics \cite {Wolf1999, Bloch1946}.
 Our approach helps to extend this geometric approach to four-level systems.

\begin{figure}
\includegraphics[width=1\columnwidth]{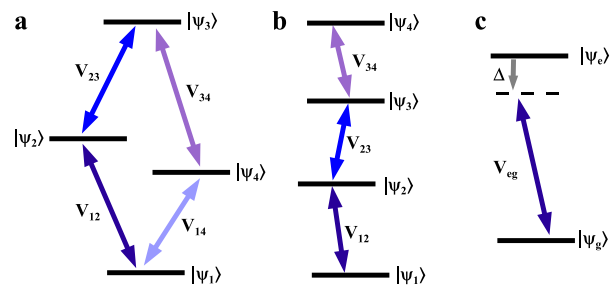}
\caption{\small Finite level realizations of the two-  and four-
mode systems of coupled equations (a) A four-level system  with
periodic nearest neighbor coupling, known as diamond shape
structure. (b) A four-level system with $V_{14}=0$, known as a
ladder type structure. (c) a simple two-level system with coupling
coefficient $V_{eg}$ and the detuning of
$\Delta$.
}\label{Fig1_Pyth_triangle}
\end{figure}

The task of finding schemes for complete population transfer between
selected states becomes increasingly difficult in  multi-mode
coupled systems. There are group-theoretical methods which provide a
rigorous tool of how to determine whether a system is
\emph{wavefunction controllable}, i.e. when any initial state in a
quantum system can be transferred into an arbitrary final state
\cite{Rabitz2001, Tannor2007}, however these methods are
nonconstructive and do not provide a general recipe on how to
implement complete population transfer scheme for a concrete system.
So far, there is only a limited number of systematic methods which
can provide this goal. The schemes exploiting adiabatic evolution
are known to be able to achieve this goal asymptotically, however
these methods require strong pair-vise sequence of coupling pulses
as well as very long dynamical time \cite{Bergmann1989,
Malinovsky1997, Vitanov1998,
Demkov1968}. 
Few solutions of the complete population transfer problem, which
requires a set of coupling coefficients to satisfy some special
relations, were found for N-mode systems \cite{Shore1979,
Eberly1992}.

In the present paper we describe a new analytical solution for a
four coupled mode system with nearest-neighbor coupling. We show a
clear similarity in the structure of this four-level solution and
the structure of the Rabi two-level solution,
originating from a common geometric character of both solutions. 
We exploit the fact that a
four-dimensional Hilbert space can be represented as a tensor
product of two two-dimensional Hilbert spaces. For the four-level nearest-neighbor coupling system the relevance of this tensor-product  structure is revealed in its full simplicity when the Hamiltonian is rewritten in the basis of the Bell states. To do a qualitative physical analysis of the resulting set of two-level equations we use
a transformation known as the Hopf map \cite{Hopf1931,
Lyons2003}. The final equations for the evolution of two separate pairs of modes
$|\psi_{1}\rangle\leftrightarrow|\psi_{3}\rangle$ and
$|\psi_{2}\rangle\leftrightarrow|\psi_{4}\rangle$ provide natural generalization of the Rabi two-mode solution for thefour-level system. Finally we derive an equation
for the time required for complete population transfer to occur.

The requirement of complete population transfer in a four-level
system  imposes certain analytical relations on the coupling
coefficients and we find that these relations have special algebraic
character: they are identical to a formula, which generates all
primitive Pythagorean triples (PPTs). Such a triple is a set of
three real numbers  $a$, $b$, $c$, which do not possess a common
factor and satisfy the equation $a^{2}+b^{2}=c^{2}$. For instance,
(3; 4; 5) and (5; 12; 13) are primitive triples, whereas (6; 8; 10)
is not PPT. Note also, that in spite of the fact that a set of
numbers (1,1,$\sqrt{2}$) satisfy the Pythagorean relation, these
numbers are \emph{not} a Pythagorean triple. Through the centuries
finding a formula to generate these triplets has intrigued both
amateur and professional mathematicians.
The first general solution was given by Euclid in his 
\emph{Elements} \cite{Euclids300BC}. It states that for any pair
$(p, q)$ of positive odd integers with $p>q$, the triple
$\left(a,b,c\right)\equiv\left(\frac{p^2-q^2}{2},pq,\frac{p^2+q^2}{2}\right)$
is Pythagorean. Other types of generating functions of PPTs can be
found elsewhere \cite{Barning1963,Arpaia1971,Wade2003}.


We start with  a nearest-neighbor coupling four-mode Hamiltonian,
when each level is coupled to two neighboring levels, as in the diamond-shaped structure, shown in Fig. \ref{Fig1_Pyth_triangle}.
\begin{equation}\label{Hamiltonian}
\hat{H}=\left(
    \begin{array}{cccc}
      0 & V_{12} & 0 & V_{14} \\
      V_{12} & 0 & V_{23} & 0 \\
      0 & V_{23} & 0 & V_{34} \\
      V_{14} & 0 & V_{34} & 0 \\
    \end{array}
  \right).
\end{equation}
As an example of physical realization
of this system, we can consider a laser-field driven four-level atom. Then in the
rotating wave approximation the coupling coefficients are
defined as
$V_{ij}=\mu_{ij}\epsilon\left(t\right)/\hbar$. 
Here $\epsilon(t)$ is the field amplitude and $\mu_{ij}$ are
 dipole matrix elements between nearest-neighbor levels $i$
and $j$. A general case of periodic nearest-neighbor
coupling (diamond-type structure), is schematically shown in Fig.
\ref{Fig1_Pyth_triangle}(a), and particular case of $V_{14}=0$,
describing a ladder-type coupling, is presented in Fig.
\ref{Fig1_Pyth_triangle}(b).

We rewrite Hamiltonian (\ref{Hamiltonian}) in the Bell basis using
the unitary transformation
\begin{equation}
\hat{W}=\frac{1}{\sqrt{2}}\left(
    \begin{array}{cccc}
      1 & 0 & 0 & 1 \\
      0 & 1 & 1 & 0 \\
      0 & 1 & -1 & 0 \\
      1 & 0 & 0 & -1 \\
    \end{array}
  \right)\:\:.
\end{equation}
The wave-functions are transformed as $\left| {\psi _n^B } \right\rangle
= W\left| {\psi _n } \right\rangle$, and the Hamiltonian becomes
$\hat{H}_{W}=\hat{W}^{\dagger}\hat{H}\hat{W}$. In this basis two
distinct $su(2)$ subalgebras, formally associated with qubits 1 and 2, can be identified. We rewrite the Hamiltonian  as a linear combination of Pauli matrices
$\hat{\sigma}_{\left(x,y,z\right)}$ acting on qubits 1 and 2
\begin{eqnarray}
\begin{array}{l}
 \hat H_W \,\, = \,\hat h^{\left( 1 \right)} \, \otimes \,\hat I^{\left( 2 \right)}  + \,\,\hat I^{\left( 1 \right)}  \otimes \hat h^{\left( 2 \right)} \: , \\
 \hat h^{\left( 1 \right)}  = \,\,\frac{{\left( {V_{23}  + V_{14} } \right)}}{2}\hat \sigma _z  + \frac{{\left( {V_{12}  - V_{34} } \right)}}{2}\hat \sigma _x \: ,  \\
 \hat h^{\left( 2 \right)}  = \,\,\frac{{\left( {V_{12}  + V_{34} } \right)}}{2}\hat \sigma _x  - \frac{{\left( {V_{23}  - V_{14} } \right)}}{2}\hat \sigma _z \: . \\
 \end{array}
\end{eqnarray} \label{WHW}

\begin{table*}
\caption{\small Comparison of the Rabi solution for two mode system
 (the middle column) with a four-mode nearest neighbor solution (the right column). We observe A striking similarity in the structure of both solutions.}
\begin{tabular}{|c|c|c|}
  \hline
  \textbf{Parameter} & \textbf{Two-mode dynamics} & \textbf{Nearest-neighbor four mode dynamics} \\
  \hline 
  Dynamical symmetry & $SU(2)$ & $SU(2)\times{SU}(2)$ \\
  Spanned space & $\{|\psi_{g}\rangle\:,|\psi_{e}\rangle\}$ & $\{|\psi_{1}\rangle\:,|\psi_{3}\rangle\}$ and
  $\{|\psi_{2}\rangle\:,|\psi_{4}\rangle\}$ \\
  Generalized frequencies & $V=\sqrt{V_{12}^2+\Delta^{2}}$ &
  $V_{L}=\sqrt{\left(V_{12}-V_{34}\right)^{2}+\left(V_{23}+V_{14}\right)^{2}}$\\
 {} & {} &
 $V_{R}=\sqrt{\left(V_{12}+V_{34}\right)^{2}+\left(V_{23}-V_{14}\right)^{2}}$\\
 "Torque" vector &
$\Omega_{R}=\left(Re\{V_{12}\},Im\{V_{12}\},\Delta\right)$ &
$\Omega_{P}=\frac{1}{\sqrt{\xi_{0}}}\left(\xi_{1},\xi_{2},\xi_{3}\right)$\\
  Ground state evolution & $a_{g}\left(t\right)=\cos\left({V{t}}\right)-\frac{\Delta}{\sqrt{V_{12}^{2}+\Delta^{2}}}\sin\left({V{t}}\right)$ & $a_{1}\left(t\right)=\cos\left(V_{L}{t}\right)\cos\left(V_{R}{t}\right)-\frac{\xi_{3}}{\sqrt{\xi_{1}^{2}+\xi_{3}^{2}}}\sin\left(V_{L}{t}\right)\sin\left(V_{R}{t}\right)$ \\
  Excited state evolution & $a_{e}\left(t\right)=-\frac{V_{12}}{\sqrt{V_{12}^{2}+\Delta^{2}}}\sin\left({V{t}}\right)$ & $a_{3}\left(t\right)=-\frac{\xi_{1}}{\sqrt{\xi_{1}^{2}+\xi_{3}^{2}}}\sin\left({V_{L}{t}}\right)\sin\left(V_{R}{t}\right)$ \\
Inversion time & $\tau=\frac{\pi}{|\Omega_{R}|}$ &
$\tau=\frac{\pi}{|\Omega_{P}|}$\\
  \hline
\end{tabular}
\end{table*}

In the language of Lie group theory Hamiltonian (\ref{Hamiltonian})
generates a subgroup $SU(2)\otimes{SU(2)}/Z_{2}$ of the full $SU(4)$
group. The dynamic problem factorizes into two separate problems for
two $SU(2)$ unitary operators acting on two qubits, thereby
geometric tools for visualization of resulting solutions are readily
available. By using an algebraic property of local transformations,
we can represent the action of an $SU(2)\times{SU(2)}/Z_{2}$
operator on a four-dimensional state vector as left and right
multiplication by two $2 \times 2$ $SU(2)$ matrices, acting on a $2
\times 2$ complex matrix. The latter represents an element of the
four-dimensional Hilbert space \cite{Piter2007}. Thus we rewrite the
equations for the evolution of amplitudes $a_{n}\left(t\right)$ of
the states $\left| {\psi _n} \right\rangle$, $n\in\{1,2,3,4\}$ in a
form of two rotations: $u_1$ acts from the left, and $u_2$ acts from
the right:
\begin{equation} \label{A-Equation}
\hat{A}\left(t\right)=\hat{u}_{1}\left(t\right)\hat{A}\left(t=0\right)\hat{u}_{2}^T\left(t\right).
\end{equation}
Here $\hat{A}\left(t\right)$ contains the information about four
amplitudes $a_{1,2,3,4}$,
\begin{equation} 
\hat A\left(t\right)  = a_1 \left( t \right)\hat I + a_2 \left( t
\right)\hat \sigma _x  + ia_3 \left( t \right) \hat \sigma _y  + a_4
\left( t \right)\hat \sigma _z \:,
\end{equation}
and the operators $u_{1}$ and $u_{2}$ are local rotations of
qubits 1 and 2 correspondingly
\begin{eqnarray*}
\hat{u}_{1}\left(t\right)&=&exp\left\{i\frac{t}{2}\left[\left(V_{12}-V_{34}\right){}\hat{\sigma}_{x}+\left(V_{23}+V_{14}\right){}\hat{\sigma}_{z}\right]\right\} \:\:, \nonumber\\
\hat{u}_{2}\left(t\right)&=&exp\left\{i\frac{t}{2}\left[\left(V_{12}+V_{34}\right){}\hat{\sigma}_{x}-\left(V_{23}-V_{14}\right){}\hat{\sigma}_{z}\right]\right\}\:\:.
\nonumber
\end{eqnarray*}\label{u1u2}
It immediately follows from this relation that if the system is initialized
in the ground state $|\psi_{1}\rangle$ so that
$a_{1}\left(0\right)=1$ and $a_{2,3,4}\left(0\right)=0$, the
amplitudes of $a_{1,3}\left(t\right)$ will remain real  and
the amplitudes of $a_{2,4}\left(t\right)$ will be purely imaginary.

To calculate explicitly $a_{n}\left(t\right)$ and analyze the
problem of complete population transfer between the states
$|\psi_{1}\rangle$ and $|\psi_{3}\rangle$, we first note that there
is another symmetry, 
which
allows us to solve the problem in an elegant geometric fashion. Suppose
that we chose to rotate the basis in the space spanned by vectors
$|\psi_{2}\rangle$ and $|\psi_{4}\rangle$. Such rotation apparently
will have no effect on the evolution of the states
$|\psi_{1}\rangle$ and $|\psi_{3}\rangle$. 
We can represent this transformation as a phase multiplication
acting on two complex vectors
$\left(V_{12}+iV_{14}\right)\rightarrow\left(V_{12}+iV_{14}\right)e^{i\theta}$
and
$\left(V_{23}+iV_{34}\right)\rightarrow\left(V_{23}+iV_{34}\right)e^{i\theta}$.
The invariance of the amplitudes $a_{1}\left(t\right)$ and
$a_{3}\left(t\right)$ under such a transformation means that the
amplitudes $a_{1}$ and $a_{3}$ are  determined not by the full set
of coupling coefficients
$\{V_{12},V_{23},V_{34},V_{14}\}\in{R^{4}}$, but by an element of
the quotient space ${R}^{2}\times R^{2}/SO(2)$,
described by a special algebraic transformation, known as the Hopf
$S^{3}\rightarrow{S^{2}}$ projective map \cite{Hopf1931,Lyons2003}.
In physics, the Hopf map is commonly associated with the Bloch
Sphere representation of a pure state. In our problem, the map takes
the 4-dimensional $V_{ij}$ space to a 3-dimensional $\xi_{n}$ space,
\begin{eqnarray}
\xi_{0}&=&\frac{1}{2}\left(V_{12}^{2}+V_{14}^{2}+V_{23}^{2}+V_{34}^{2}\right)\:,\nonumber\\
\xi_{1}&=&V_{12}V_{23}+V_{14}V_{34}\:,\nonumber\\
\xi_{2}&=&V_{12}V_{34}-V_{23}V_{14}\:,\nonumber\\
\xi_{3}&=&\frac{1}{2}\left(V_{12}^{2}+V_{14}^{2}-V_{23}^{2}-V_{34}^{2}\right)\:.
\end{eqnarray}
The new coordinates $\xi_{n}$ satisfy the equation
$\xi_{0}^{2}-\xi_{1}^{2}-\xi_{2}^{2}-\xi_{3}^{2}=0$, which is the
equation for a 3-dimensional cone embedded in four-dimensional
Eucledian space \cite{Kocik2006}. Physical meaning of this set of
parameters will be clarified when we use them to derive the final
solution for the amplitudes $a_n\left(t\right)$.

Time dependent amplitudes of the modes $|\psi_{1}\rangle$ and
$|\psi_{3}\rangle$ can be expressed directly from equation (\ref{A-Equation}) using coupling coefficients $V_{i,j}$, however by using the projective coordinates
$\{\xi_{1},\xi_{2},\xi_{3}\}$, algebraic expressions for these amplitudes can be significantly compactified. As it immediately follows from equation (\ref{A-Equation}),
\begin{eqnarray}\label{amplitudes13}
a_{3}\left(t\right)&=&-\frac{\xi_{1}}{\sqrt{\xi_{1}^{2}+\xi_{3}^{2}}}\sin\left({V_{L}{t}}\right)\sin\left(V_{R}{t}\right),\\
a_{1}\left(t\right)&=&\cos\left(V_{L}{t}\right)\cos\left(V_{R}{t}\right)-\frac{\xi_{3}}{\sqrt{\xi_{1}^{2}+\xi_{3}^{2}}}\sin\left(V_{L}{t}\right)\sin\left(V_{R}{t}\right). \nonumber
\end{eqnarray}
Here
$V_{L,R}\equiv\sqrt{\frac{1}{2}\left(\xi_{0}\mp\xi_{2}\right)}$ are
the left and right frequencies, respectively. Explicit formulas
for $V_{L}$ and $V_{R}$ are given in Table I.
\begin{figure}
\includegraphics[width=1\columnwidth]{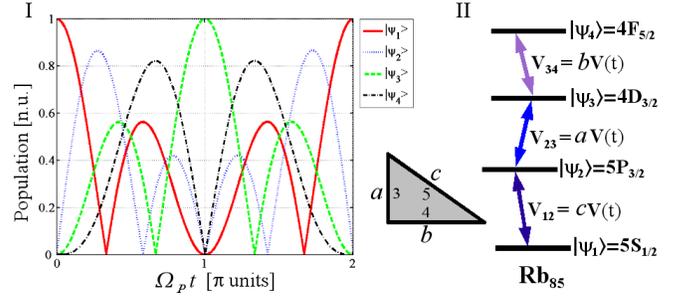}
\caption{\small Complete population transfer
$|\psi_1\rangle\longleftrightarrow|\psi_3\rangle$ (or
$|\psi_2\rangle\longleftrightarrow|\psi_4\rangle$). (I) Level
populations as functions of interaction time, for a ladder-type
coupling scheme. The system initially is prepared in the ground
state (solid red). The population is periodically transferred to the
third state (dashed green). Parameters correspond to $(p,q)=(3,1)$
in Eq. 8. (II) The Pythagorean triple relation between the coupling
coefficients in four level ladder system. The computed transition
time for complete population transfer matches the calculated
transition time from Eq. 9.} \label{Exp_4level_dim}
\end{figure}\label{Fig2}

The form of the exact solution given by equation (\ref{amplitudes13}) is similar to
the form or the Rabi solution
for a two-level system, as demonstrated in Table I. 
In the two-level case, shown in Fig. 1(c), we use notations $a_{g}$
and $a_{e}$ for the amplitudes of the ground and excited states,
while detuning was denoted as
$\Delta=(\omega_{e}-\omega_{g})-\omega_{laser}$. As can be seen from
Table I, instead of one  frequency $V$ for the two-mode dynamics,
the four-mode system is characterized by two generalized frequencies
$V_{L}$ and $V_{R}$, and instead of single sine and cosine functions
in the two mode case, we see that the four-mode dynamics is
determined by the product of two sine and two cosine functions. The
pre-factors of the sine function in both cases share a similar
structure, being formally equal to  x- and z- components of unit
vectors. One can see that there is a very close  analogy between
variables $\xi_{1,2,3}$ and components of the the torque vector for
two-mode dynamics \cite{Eberly1975}.

 Now we have all the necessary equations to solve the problem of complete population
 transfer. To realize complete population transfer from the state $|\psi_{1}\rangle$ to the state
$|\psi_{3}\rangle$ at time $t=\tau$ one has to set the $\xi_{3}$
variable equal to zero, in the same fashion as the detuning $\Delta$
in the two-mode case has to equal zero (the requirement of on
resonant interaction). Next, complete population transfer will occur
only when dynamic angles $V_{L}t$ and $V_{R}t$, simultaneously
complete a $\pi$-phase rotation, i.e. when
$V_{L}=\frac{\pi}{2\tau}\left(2m_{1}+1\right)\equiv\frac{\pi}{2\tau}p$
and
$V_{R}=\frac{\pi}{2\tau}\left(2m_{2}+1\right)\equiv\frac{\pi}{2\tau}q$.
After some trivial algebra, we derive the following solution:
\begin{eqnarray}
\left(\xi_{0},\xi_{1},\xi_{2}\right)=\frac{\pi^{2}}{2\tau^{2}}\left(\frac{{p}^2+q^2}{2},pq,\frac{{p}^2+q^2}{2}\right)=\frac{\omega^{2}}{2}\left(c,a,b\right) \: .
\end{eqnarray} \label{xi-solution}
This solution exactly matches the definition of
the generating function of primitive Pythagorean triples
\cite{Euclids300BC}. For the nearest-neighbor 4-mode coupling
problem complete population transfer can occurs between two
nonadjacent states $|\psi_{1}\rangle $ and $|\psi_{3}\rangle$ (or
$|\psi_{2}\rangle $ and $|\psi_{4}\rangle$) only when the ratio
$(\xi_{0}:\xi_{1}:\xi_{2})$ is equal to
ratio of a Pythagorean triple. 
The equation for coefficients
$\{V_{12},V_{23},V_{34},V_{14}\}$, can be obtained from equation (8) by inverting the
transformation (6). For the special case of latter-type coupling, where $V_{14}=0$,
this solution becomes $\xi_{0}=V_{12}^2$, $\xi_{1}=V_{23}V_{12}$,
$\xi_{2}=V_{34}V_{12}$ such that relation between the
nearest-neighbor coupling coefficients and the Pythagorean triple
takes a simple form of a proportion $(V_{12};V_{23};V_{34})\sim(c,a,b)$.

We tested our theoretical prediction by performing numerical
simulations on the dynamics of a four-level ladder transitions in
$Rb_{85}$: 
$5S_{1/2}\leftrightarrow{5P_{3/2}}\leftrightarrow{4D_{3/2}}\leftrightarrow{4F_{5/2}}$,
with resonant CW interaction of $780.2nm$, $1.529\mu{m}$,
$1.344\mu{m}$, respectively. The coupling coefficients were chosen
to satisfy the simplest Pythagorean triple ratio
$(V_{12}:V_{23}:V_{34})\sim(5:3:4)$. As seen in Fig. 2, numerical
results are in complete agreement with the  analytical solution and
confirm that there is periodic population transfer between states
$|\psi_{1}\rangle$ and $|\psi_{3}\rangle$. The time period for
complete population transfer is given by
\begin{equation}
\tau\equiv\frac{\pi}{\Omega_{P}}=\frac{\pi}{\sqrt{V_{L}^{2}+V_{R}^{2}}}=\frac{\pi}{\sqrt{2\xi_{0}}}\:\:.
\end{equation}\label{transition_time}
Here we denoted the transition time required to achieve
population transfer as $\Omega_{P}$, 
analogous with the Rabi frequency for a two-level
dynamics. Note that this parameter scales as the absolute value of
the torque
vector, similar with the two-mode case.

In conclusion, we identified a new scheme for complete population transfer
in a four-mode systems. We observed very close connection
between the structure of solution for the nearest-neighbor
coupling four-level system and the generating function of
primitive Pythagorean triples.
This solution can be used not only for time-dependent problems, but
also for some problems of spatial propagation of light pulses, such
as coupling between directional waveguides. We expect that similar
solutions, revealing deeper link with the number theory, can be
found for 6- and  8-level systems. The present method, describing
the four-level dynamics, can be generalized to include more complex
exactly solvable two-level models. This work is in progress.

This research was supported by ISF and by the NSF under Grants PHY-0545390 and in part by the National Science Foundation under Grant No. PHY05-51164 for the Kalvi Institute for
 Theoretical Physics, UCSB. One of us (HS) is grateful to the Azrieli Foundation for financial
 support.


\begin{thebibliography}{25}
\expandafter\ifx\csname
natexlab\endcsname\relax\def\natexlab#1{#1}\fi
\expandafter\ifx\csname bibnamefont\endcsname\relax
  \def\bibnamefont#1{#1}\fi
\expandafter\ifx\csname bibfnamefont\endcsname\relax
  \def\bibfnamefont#1{#1}\fi
\expandafter\ifx\csname citenamefont\endcsname\relax
  \def\citenamefont#1{#1}\fi
\expandafter\ifx\csname url\endcsname\relax
  \def\url#1{\texttt{#1}}\fi
\expandafter\ifx\csname urlprefix\endcsname\relax\def\urlprefix{URL
}\fi \providecommand{\bibinfo}[2]{#2}
\providecommand{\eprint}[2][]{\url{#2}}

\bibitem[{\citenamefont{Allen and Eberly}(1975)}]{Eberly1975}
\bibinfo{author}{\bibfnamefont{L.~D.} \bibnamefont{Allen}} \bibnamefont{and}
  \bibinfo{author}{\bibfnamefont{J.~H.} \bibnamefont{Eberly}}
  (\bibinfo{publisher}{Wiley, New York}, \bibinfo{year}{1975}).

\bibitem[{\citenamefont{Keeler, J.}(2005)}]{NMR2005}
\bibinfo{author}{\bibfnamefont{J.} \bibnamefont{Keeler}}
  (\bibinfo{publisher}{John Wiley and Sons}, \bibinfo{year}{2005}).

\bibitem[{\citenamefont{Nielsen and Chuang}(2000)}]{Nielsen2000}
\bibinfo{author}{\bibfnamefont{M.~A.} \bibnamefont{Nielsen}} \bibnamefont{and}
  \bibinfo{author}{\bibfnamefont{I.~L.} \bibnamefont{Chuang}}
  (\bibinfo{publisher}{Cambridge University Press, Cambridge},
  \bibinfo{year}{2000}).

\bibitem[{\citenamefont{{Sorenson} and Klaus}(1999)}]{Sorenson1999}
\bibinfo{author}{\bibfnamefont{A.}~\bibnamefont{{Sorenson}}} \bibnamefont{and}
  \bibinfo{author}{\bibfnamefont{M.}~\bibnamefont{Klaus}}
  (\bibinfo{year}{1999}).

\bibitem[{\citenamefont{{Yariv}}(1973)}]{Yariv1973}
\bibinfo{author}{\bibfnamefont{A.}~\bibnamefont{{Yariv}}},
  \bibinfo{journal}{Phys. Rev.} \textbf{\bibinfo{volume}{70}},
  \bibinfo{pages}{460} (\bibinfo{year}{1973}).

\bibitem[{\citenamefont{{Rau} and {Uskov}}(2000)}]{Rau2000}
\bibinfo{author}{\bibfnamefont{A.~R.} \bibnamefont{{Rau}}} \bibnamefont{and}
  \bibinfo{author}{\bibfnamefont{D.}~\bibnamefont{{Uskov}}},
  \bibinfo{journal}{Phys. Rev. A} \textbf{\bibinfo{volume}{61}},
  \bibinfo{pages}{032301} (\bibinfo{year}{2000}).

\bibitem[{\citenamefont{{Uskov} and {Rau}}(2008)}]{Uskov2008}
\bibinfo{author}{\bibfnamefont{D.}~\bibnamefont{{Uskov}}} \bibnamefont{and}
  \bibinfo{author}{\bibfnamefont{A.~R.} \bibnamefont{{Rau}}},
  \bibinfo{journal}{Phys. Rev. A} \textbf{\bibinfo{volume}{78}},
  \bibinfo{pages}{022331} (\bibinfo{year}{2008}).

\bibitem[{\citenamefont{{Born} and {Wolf}}(1999)}]{Wolf1999}
\bibinfo{author}{\bibfnamefont{M.}~\bibnamefont{{Born}}} \bibnamefont{and}
  \bibinfo{author}{\bibfnamefont{E.}~\bibnamefont{{Wolf}}}
  (\bibinfo{publisher}{Cambridge University Press}, \bibinfo{year}{1999}).

\bibitem[{\citenamefont{{Bloch}}(1946)}]{Bloch1946}
\bibinfo{author}{\bibfnamefont{F.}~\bibnamefont{{Bloch}}},
  \bibinfo{journal}{Phys. Rev.} \textbf{\bibinfo{volume}{70}},
  \bibinfo{pages}{460} (\bibinfo{year}{1946}).

\bibitem[{\citenamefont{{Rabitz} and {Turinici}}(2001)}]{Rabitz2001}
\bibinfo{author}{\bibfnamefont{H.}~\bibnamefont{{Rabitz}}} \bibnamefont{and}
  \bibinfo{author}{\bibfnamefont{G.}~\bibnamefont{{Turinici}}},
  \bibinfo{journal}{Chemical Physics} \textbf{\bibinfo{volume}{267}},
  \bibinfo{pages}{1} (\bibinfo{year}{2001}).

\bibitem[{\citenamefont{{Tannor}}(2007)}]{Tannor2007}
\bibinfo{author}{\bibfnamefont{D.}~\bibnamefont{{Tannor}}}
  (\bibinfo{publisher}{University Science Books}, \bibinfo{year}{2007}).

\bibitem[{\citenamefont{{Kuklinski} et~al.}(1989)\citenamefont{{Kuklinski},
  {Gaubatz}, {Hioe}, and {Bergmann}}}]{Bergmann1989}
\bibinfo{author}{\bibfnamefont{J.~R.} \bibnamefont{{Kuklinski}}},
  \bibinfo{author}{\bibfnamefont{U.}~\bibnamefont{{Gaubatz}}},
  \bibinfo{author}{\bibfnamefont{F.~T.} \bibnamefont{{Hioe}}},
  \bibnamefont{and}
  \bibinfo{author}{\bibfnamefont{K.}~\bibnamefont{{Bergmann}}},
  \bibinfo{journal}{Phys. Rev. A} \textbf{\bibinfo{volume}{40}},
  \bibinfo{pages}{6741 } (\bibinfo{year}{1989}).

\bibitem[{\citenamefont{{Malinovsky} and {Tannor}}(1998)}]{Malinovsky1997}
\bibinfo{author}{\bibfnamefont{V.~S.} \bibnamefont{{Malinovsky}}}
  \bibnamefont{and} \bibinfo{author}{\bibfnamefont{D.~J.}
  \bibnamefont{{Tannor}}} (\bibinfo{year}{1998}).

\bibitem[{\citenamefont{{Vitanov} et~al.}(1998)\citenamefont{{Vitanov},
  {Shore}, and {Bergmann}}}]{Vitanov1998}
\bibinfo{author}{\bibfnamefont{N.~V.} \bibnamefont{{Vitanov}}},
  \bibinfo{author}{\bibfnamefont{B.~W.} \bibnamefont{{Shore}}},
  \bibnamefont{and}
  \bibinfo{author}{\bibfnamefont{K.}~\bibnamefont{{Bergmann}}},
  \bibinfo{journal}{Eur. Phys. J. D} \textbf{\bibinfo{volume}{4}},
  \bibinfo{pages}{29} (\bibinfo{year}{1998}).

\bibitem[{\citenamefont{{Demkov} and {Osherov}}(1968)}]{Demkov1968}
\bibinfo{author}{\bibfnamefont{Y.~N.} \bibnamefont{{Demkov}}} \bibnamefont{and}
  \bibinfo{author}{\bibfnamefont{V.~I.} \bibnamefont{{Osherov}}}
  (\bibinfo{year}{1968}).

\bibitem[{\citenamefont{{Shore} and {Cook}}(1979)}]{Shore1979}
\bibinfo{author}{\bibfnamefont{B.}~\bibnamefont{{Shore}}} \bibnamefont{and}
  \bibinfo{author}{\bibfnamefont{K.}~\bibnamefont{{Cook}}},
  \bibinfo{journal}{Phys. Rev.}  (\bibinfo{year}{1979}).

\bibitem[{\citenamefont{{Shore} et~al.}(1992)\citenamefont{{Shore}, {Bergmann},
  {Kuhn}, {Sciemann}, {Oreg}, and {Eberly}}}]{Eberly1992}
\bibinfo{author}{\bibfnamefont{B.}~\bibnamefont{{Shore}}},
  \bibinfo{author}{\bibfnamefont{K.}~\bibnamefont{{Bergmann}}},
  \bibinfo{author}{\bibfnamefont{A.}~\bibnamefont{{Kuhn}}},
  \bibinfo{author}{\bibfnamefont{S.}~\bibnamefont{{Sciemann}}},
  \bibinfo{author}{\bibfnamefont{J.}~\bibnamefont{{Oreg}}}, \bibnamefont{and}
  \bibinfo{author}{\bibfnamefont{J.~H.} \bibnamefont{{Eberly}}}
  (\bibinfo{year}{1992}).

\bibitem[{\citenamefont{{Hopf}}(1931)}]{Hopf1931}
\bibinfo{author}{\bibfnamefont{H.}~\bibnamefont{{Hopf}}},
  \bibinfo{journal}{Mathematische Annalen} \textbf{\bibinfo{volume}{104}},
  \bibinfo{pages}{637–665} (\bibinfo{year}{1931}).

\bibitem[{\citenamefont{{Lyons}}(2003)}]{Lyons2003}
\bibinfo{author}{\bibfnamefont{D.~W.} \bibnamefont{{Lyons}}},
  \bibinfo{journal}{Mathematics Magazine} \textbf{\bibinfo{volume}{76}},
  \bibinfo{pages}{87} (\bibinfo{year}{2003}).

\bibitem[{Euc(2007)}]{Euclids300BC}
\bibinfo{journal}{http://aleph0.clarku.edu/~djoyce/java/elements/elements.html}
   (\bibinfo{year}{2007}).

\bibitem[{\citenamefont{{Barning}}(1963)}]{Barning1963}
\bibinfo{author}{\bibfnamefont{F.~J.~M.} \bibnamefont{{Barning}}},
  \bibinfo{journal}{Math. Centrum Amsterdam Afd. Zuivere Wisk}
  p.~\bibinfo{pages}{37} (\bibinfo{year}{1963}).

\bibitem[{\citenamefont{{Arpaia}}(1971)}]{Arpaia1971}
\bibinfo{author}{\bibfnamefont{P.~J.} \bibnamefont{{Arpaia}}},
  \bibinfo{journal}{Mathematics Magazine} \textbf{\bibinfo{volume}{44}},
  \bibinfo{pages}{26} (\bibinfo{year}{1971}).

\bibitem[{\citenamefont{{McCullough} and {Wade}}(2003)}]{Wade2003}
\bibinfo{author}{\bibfnamefont{D.}~\bibnamefont{{McCullough}}}
  \bibnamefont{and} \bibinfo{author}{\bibfnamefont{E.}~\bibnamefont{{Wade}}},
  \bibinfo{journal}{The College Mathematics Journal}
  \textbf{\bibinfo{volume}{34}}, \bibinfo{pages}{107} (\bibinfo{year}{2003}).

\bibitem[{\citenamefont{{Piter}}(2007)}]{Piter2007}
\bibinfo{author}{\bibnamefont{{Piter}}}, \bibinfo{journal}{Chemical Physics}
  \textbf{\bibinfo{volume}{167}}, \bibinfo{pages}{2} (\bibinfo{year}{2007}).

\bibitem[{\citenamefont{{Kocik}}()}]{Kocik2006}
\bibinfo{author}{\bibfnamefont{J.}~\bibnamefont{{Kocik}}}, \bibinfo{journal}{Adv. appl. Clifford alg.}
  \textbf{\bibinfo{volume}{17}}, \bibinfo{pages}{71-93} (\bibinfo{year}{2006}).

\end{thebibliography}

\end{document}